\documentclass[aip,floatfix,amsmath,amssymb,reprint,superscriptaddress]{revtex4-2}

\usepackage{graphicx}% Include figure files
\usepackage{dcolumn}% Align table columns on decimal point
\usepackage{bm}
\usepackage[colorlinks=True,allcolors=blue]{hyperref}
\usepackage{newtxtext,newtxmath}
\usepackage{microtype}
\usepackage{lipsum}
\usepackage{amsmath,amssymb}
\usepackage{pdfpages}
\usepackage{sidecap}
\usepackage[detect-all]{siunitx}

\makeatletter %This fixes pdfpages rotating everything
\AtBeginDocument{\let\LS@rot\@undefined}
\makeatother
\makeatletter
\renewcommand\@make@capt@title[2]{%
 \@ifx@empty\float@link{\@firstofone}{\expandafter\href\expandafter{\float@link}}%
  {\textbf{#1}}\@caption@fignum@sep#2\quad
}%
\makeatother

\newcommand{\tud}{Department of Chemical Engineering, Delft University of
Technology, Delft 2629 HZ, The Netherlands}
\newcommand{\ue}{Currently at: SUPA and School of Physics and Astronomy, The University of Edinburgh, Peter Guthrie Tait Road, Edinburgh EH9 3FD, United Kingdom}
\newcommand{\msc}{Laboratoire Mati\`ere et Syst\`emes Complexes, CNRS UMR 7057, Universit\'e Paris Cit\'e, Paris, France}
\newcommand{\ens}{Department of Chemistry, ENS Paris, 24 rue Lhomond, 75005 Paris, France}
\newcommand{\upenn}{Department of Chemical and Biomolecular Engineering, University of Pennsylvania, 220 South 33rd Street, Philadelphia, PA 19104-6393, United States}

\begin{document}

\title{Microstructural features governing fracture of a two-dimensional amorphous solid identified by machine learning}

\author{Max Huisman}
 \email{m.huisman@sms.ed.ac.uk}
 \affiliation{\tud}
 \affiliation{\ue}
\author{Axel Huerre}
\affiliation{\msc}
\author{Saikat Saha}
\affiliation{\ens}
\author{John C.\ Crocker}
\affiliation{\upenn}
\author{Valeria Garbin}
\email{v.garbin@tudelft.nl}
    \affiliation{\ue}

\begin{abstract}
    Brittle fracturing of materials is common in natural and industrial processes over a variety of length scales. Knowledge of individual particle dynamics is vital to obtain deeper insight into the atomistic processes governing crack propagation in such materials, yet it is challenging to obtain these details in experiments. We propose an experimental approach where isotropic dilational strain is applied to a densely packed monolayer of attractive colloidal microspheres, resulting in fracture. Using brightfield microscopy and particle tracking, we examine the microstructural evolution of the monolayer during fracturing. Furthermore, using a quantified representation of the microstructure in combination with a machine learning algorithm, we calculate the likelihood of regions of the monolayer to be on a crack line, which we term \textit{Weakness}. From this analysis, we identify the most important contributions to crack propagation and find that local density is more important than orientational order. Our methodology and results provide a basis for further research on microscopic processes during the fracturing process.
\end{abstract}

\maketitle

%%%MAIN TEXT%%%%

\section{Introduction}

Cracks occur over natural length scales from atoms to earthquakes, but a thorough understanding remains elusive due the unpredictable nature of the fracture process. Generally, materials that fracture under sufficiently high strain are referred to as  \textit{brittle}. Brittle materials display a discontinuous drop in stress, in contrast with the continuous evolution over strain of ductile materials. 

The field of fracture mechanics was revolutionized by seminal work of A.A. Griffiths \cite{Griffith1920}, showing how the decrease of the strain energy by breaking the particle bonds should be higher than the increase in surface energy due to the formation of the free surface during fracturing. These results were generalized to any ``somewhat brittle'' material in later work \cite{Irwin1957}, in which also the main failure modes during fracturing were identified: shear cracks form when stress is applied parallel to the plane of the crack, whereas extensional cracks form when a tensile stress is applied normal to the plane of the crack. Other important early findings show how the stress distribution changes around the propagating crack front \cite{Williams1961, Westergaard1939}. The fracture mechanics theories from these reports use a continuum description of the material, causing the theory to break down near the crack's tip, where the stress field diverges \cite{Rountree2002}. Since the processes occurring in vicinity of the crack tip are vital in determining the macroscopic process of crack growth and propagation through a material \cite{Rozen-Levy2020}, it is important to study the dynamics at the small scale. \par

Recent advancements in simulations and experimental methods have accelerated research into the dynamic material evolution near the crack tip. In simulations, it was shown that cracks tend to initiate in the regions with highest disorder of a brittle amorphous material \cite{ozawa2021rare} and that the direction of crack propagation can be substantially influenced by the presence of defects and voids that lie in front of the crack tip \cite{Rountree2002, Kalia1997}. These findings were confirmed in experiments where the dynamic fracturing of brittle polymeric gels was studied using optical microscopy, showing the important role of defects and voids in crack propagation \cite{Ravi-Chandar1984,Rozen-Levy2020}. \par

Observations from simulations \cite{ozawa2021rare,Pollard2022} and scattering experiments \cite{Aime2018} strengthen this view by showing how fracturing is governed by localized plastic rearrangements of individual particles, which occur in ``soft regions''. Soft regions are regions in a material where particles are most likely to rearrange, characterized by low density and/or high disorder. In the case of attractive particles, particles in soft regions have fewer neighbours that fix their position. Experimental observations on individual particle dynamics in such soft regions during fracturing would be crucial for obtaining a better understanding of the role of microstructure during fracturing, but have to this date not been reported. Individual particle dynamic are often studied using small colloidal particles sized $\sim 100~$nm-$10~\mu$m, due to ease of use in combination with various optical microscopy techniques.

%We propose that experimental systems utilizing small colloidal particles can provide an excellent basis for such experiments.  

Related to fracturing, colloidal systems with small ($<100~$nm) particles have been used to study macroscale fracturing during drying, relevant to, for instance, the aging of paintings \cite{Giorgiutti2016} or dairy stratification \cite{Floch2019}. To allow for live tracking of individual particle movements, colloidal systems with larger particles of size $\sim 1~\mu$m should be used. When using a monolayer of such colloids on an interface and applying a strain to that monolayer, movements are enforced, which potentially leads to yielding. Previously, such experimental systems have been used to study among others the role of defects \cite{Buttinoni2017a}, the relaxation time scaling in plastic flow under oscillatory shear \cite{Galloway2020}, and the impact propagation through a monolayer after a localized mechanical pulse \cite{Buttinoni2017}.   

One of the main advantages of individual particle tracking is that one can quantify the microstructure from the particle coordinates over time, through so-called ``structural indicators'' \cite{Richard2020}. These parameters characterize some important features of the system, such as the local density (for instance through the number of nearest neighbours, or the area of cells in a Voronoi tessellation) or the local order (for instance through orientational order parameters $\psi_{i}$). Recently, it has been suggested that simple, machine learning (ML) algorithms can also be used to predict how likely individual particles in a sheared system are to undergo plastic rearrangement \cite{Schoenholz2016,Cubuk2015,Sharp10943}. The structural indicator called \textit{Softness} characterizes the local structure and was found to be strongly linked to the system dynamics \cite{Schoenholz2016}. An extra incentive for applying ML algorithms to such experimental systems, is that it can be used to identify the most important features in the provided dataset through analyzing the decision making process.

In this paper, we test the extension of such machine learning based methods to experimental systems with macroscopic catastrophic yielding, like fracturing. First, we develop an experimental method where a monolayer of attractive colloids is fractured by applying an isotropic strain. Using brightfield microscopy and particle tracking algorithms we extract particle coordinates, that we use to characterize the monolayer structure and its dynamic evolution. This is done by calculating the orientational bond order parameter and number of nearest neighbours. Since fracture nucleation is a stochastic process, we extend our analyses by using a machine learning method \cite{Schoenholz2016,Cubuk2015,Sharp10943} to \textit{a priori} identify regions that are more likely to be on a crack line than others, and we term this structural likelihood the \textit{Weakness}. Finally, we obtain a deeper understanding into the fracturing process by comparing the relative importance of the input features of the machine learning algorithm.

\section{Materials and Methods}

\begin{figure*}[t!]
 \centering
 \includegraphics[width=0.8\textwidth]{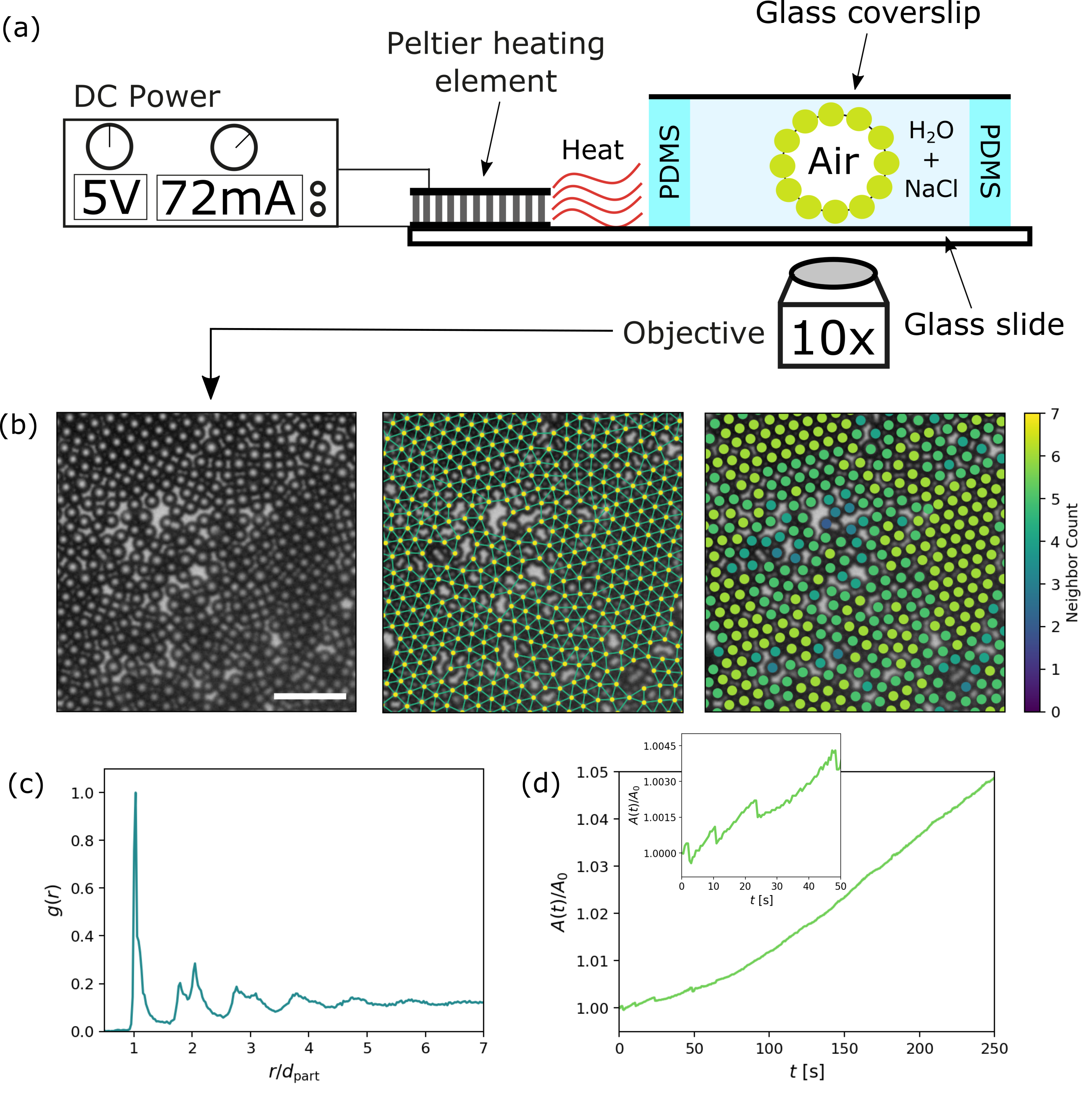}
 \caption{\textbf{(a)} Experimental schematic of the container for a colloid coated air bubble in water. Fracturing is induced through a Peltier heating element connected to DC power, and data is obtained through an optical microscope connected to a camera. \textbf{(b)} A segment of the colloid monolayer, visualized through the raw data, the Delauney Triangulation and the amount of Nearest Neighbours. Scalebar represents $25~\mu$m. \textbf{(c)} Radial distribution function $g(r)$ for a typical experimental dataset ($\sim5000~$particles). $g(r)$ is normalized by the first peak value and the radial distance from the central particle $r$ is normalized by the measured effective distance between particles $d_{\rm{part}}=r_{ij}=5.4~\mu$m. \textbf{(d)} Areal expansion ($A(t)$) of the perimeter of a colloid coated air bubble in water, normalized by it's initial value $A_0$. This data was obtained using $5\times$ magnification to track the entire perimeter over time. Inset shows a zoom-in of the initial stages of areal expansion.}
 \label{fig1}
\end{figure*}

\subsection{Sample preparation}
\label{sec:sample_preparation}

We use polystyrene microspheres with negatively charged sulfate functional groups (nominal $d_{\rm{avg}} = 5 \pm 0.5~ \mu$m, ThermoFisher, cat. number: S37227, material lot number: 853189). The colloid suspension ($4\%$w/v) was washed repeatedly by centrifugating and replacing the supernatant with milli-\textit{Q} water to remove possible contaminations. The suspension was diluted to $0.4\%$w/v using a $500~$mM NaCl aqueous solution to screen electrostatic repulsion between particles and promote adsorption to the gas-water interface. 

To produce colloid-coated air bubbles in water, we thoroughly shake the colloidal suspension to create air bubbles, which also agitates the colloids so that they adsorb at the interfaces of the air bubbles in water. The resulting colloid-coated bubbles are sufficiently stable that they can be individually extracted from the vial using a spatula. The bubble was then placed atop a sample holder, consisting of a $4~$mm thick PDMS spacer on a glass slide ($76 \times 26~$ mm$^{2}$), filled with a $500~$mM NaCl solution and subsequently covered by a glass coverslip ($18 \times 18~$ mm$^{2}$). Next, the sample was left undisturbed for at least 10 minutes to equilibrate. A Peltier heating element (RS Peltier Module, $1.6~$W, $1.6~$A, $7~$V, $30 \times 30~$ mm$^{2}$) was glued close to the container, for controlling the temperature of the sample. After preparation, the entire sample container was placed on an inverted microscope (IX71, Olympus) equipped with a camera (Basler ace acA5472-17uc) and a $10\times$ objective. A schematic of the experimental setup is shown in Fig. \ref{fig1} (a).

The in-focus part of the monolayer in the field of view [see Fig.~\ref{fig1} (b)] contains $\sim 5000~$ particles. The typical surface coverage is $\phi=\frac{N_{\rm{part}}\pi R_{\rm{part}}^{2}}{A_{\rm{surf}}} = 0.72 \pm 0.05$, $N_{\rm{part}}$ being the amount of particles in the field of view, $R_{\rm{part}}$ the particle radius and $A_{\rm{surf}}$ the area of the field of view containing in-focus particles. We note that in our experiments, we found an effective center-to-center distance of $r_{ij} = R_{\rm{part}}/2 \approx 5.4~\mu$m between particles $i$ and $j$.

\subsection{Controlled monolayer expansion and fracture}
\label{sec:heating}

In previous work, dynamics of colloidal monolayers under strain have for instance been studied through the inflation/deflation of a pendant drop \cite{Huerre2018} or by cooling-induced shrinkage of compressible air droplets in water \cite{Poulichet2015_langmuir}. Here, we heat colloid coated air bubbles to study extensional fracturing while tracking individual particles, which would otherwise be difficult to study simultaneously. 

We expand air bubbles in water by heating the sample holder using a Peltier element. The fluid in the sample holder was heated only by a couple of degrees, ensuring slow expansion to allow for particle tracking. 

The areal expansion of the perimeter of a colloid-coated bubble in a typical experiment is shown in Fig. \ref{fig1} (d). The growth rate slowly increases for $t <100~$s. During this stage also some sudden drops can be observed (inset of \ref{fig1} (d)), possibly indicating rapid changes in the structure of the monolayer. After this initial stage ($t > 100~$s), the bubble area increases at a constant rate. We find possible explanations for this behaviour by zooming in at the evolution of a monolayer over time (see Fig. \ref{fig2}(a)), where we see that the new crack formation mainly occurs in the early stages of bubble growth. These results show a rapid crack propagation through the monolayer that seems heavily influenced by the orientation of the initial crack (more examples of the crack directionality can be observed in Fig. S1 in the Supplemental Information). The rapid changes in the monolayer structure at early times could result in the abrupt changes we observed in the measured perimeter of the bubble in the inset of Fig. \ref{fig1} (d). Next, the bubble proceeds to grow through areal expansion of the already formed cracks, rather than new crack formation, which we assume corresponds to the constant growth rate of the bubble at later times. \par

\subsection{Image analysis and particle tracking}

To obtain particle coordinates from our microscopy data we used manual particle tracking algorithms in Matlab. We found that a completely automated approach was insufficiently accurate as our microscopy data contains many ($\sim 5000~$) particles that are subjected to sudden movement, move slightly in and out of focus during an experiment and where the difference between a void and a particle is difficult to detect.

In our approach, we first obtain estimates of initial coordinates using TrackPy \cite{dan_allan_2019_3492186}. These were imported to Matlab and refined by manually removing voids classified as particles and adding particles that were not recognised. Particle tracking was performed using the Crocker and Grier algorithm\cite{CROCKER1996298}. When particles experienced a sudden rapid movement or when they moved out of focus such that the tracking algorithm lost a particle, we re-adjusted this particle's position by hand. With an image resolution of 150 nm per pixel, particle tracking has a subpixel accuracy. To make sure we had enough data to use machine learning algorithms, we performed 20 identical experiments, which combined together form a dataset containing trajectories of approximately 100,000 particles.

\subsection{Fracture detection}

To quantify the crack location in the monolayer, we identify the particles located on the boundary of a crack. To this end we adapted an image analysis routine originally developed to visualize pore connectivity in metal-organic frameworks \cite{mayorga2021visualizing}. We converted the final frame of each experiment to a binary mask, that only contained cracks with a total size above a minimum pixel size, that we adjusted manually for each experiment. This mask was morphologically dilated by 1 particle diameter and overlaid onto the particle coordinates in the final frame to identify the particles on the edge of a crack. In a typical experiment, a subset of about 100 to 200 particles were located on the edges of cracks out of $\sim 5000$ total particles in the field of view.

\subsection{Quantifying the local microstructure}

A segment of a typical experimental monolayer is shown in Fig. \ref{fig1} (b). We can observe rafts of localized crystalline order, where particles have 6 nearest neighbours (e.g. a hexagonal centered packing arrangement), interchanged by more amorphous regions. This observation is in close resemblance to systems from other studies on colloid monolayers with medium range ordering \cite{Schwenke2014,Keim2014,Galloway2020}. The radial distribution function $g(r)$ (Fig. \ref{fig1} (c)) confirms this similarity, with local ordering extending up to about 6 coordination shells.  %refer to surface coverage, NN etc. \par

We characterize and investigate the monolayer structure using the number of nearest-neighbours (NN), which relates to the local density, and the bond order parameter $\Psi_{i}$, relating to the local orientational order. NN is the number of particles within a cut-off radius $R_{\rm{c}}=2*r_{\rm{part}}$ ($=r_{ij}=5.4~\mu$m). We calculate the hexatic bond order parameter as \cite{Richard2020}

\begin{equation}
    \psi_6^{i}=\frac{1}{n_{i}} \left | \sum^{n_{i}}_{j}e^{i6\theta_{ijk}} \right |
\end{equation}

where $i$ is the central particle of interest, $\theta_{ijk}$ is the angle of particle $i$ with neighbours $j$ and $k$. Note that $i$ in the exponent is the unit imaginary number. $\psi_6^{i}$ is a measure of hexagonal order, with $\psi_6^{i} \rightarrow 1$ for perfect hexagonal order and $\psi_6^{i} \rightarrow 0$ otherwise. 

\subsection{A generalized description of local microstructure}
\label{sec:gen_parameters}

We also calculate a structural indicator from a more general description of the local particle environment, using machine learning algorithms. This approach was first proposed by Behler and Parinello \cite{Behler2007} and later applied to the study of plastic rearrangements in colloidal systems \cite{Schoenholz2016,Cubuk2015,Sharp10943}.

The generalized description consists of two \textit{structure functions}. The first structure function $G_{Y}^{X} (i,\mu)$ essentially acts as a discretized radial distribution function, that calculates how many neighbours $j$ are located in a shell of thickness $\delta$ at a distance $\mu$ from particle $i$, and is defined as

\begin{equation}
    \label{eq:G_func}
    G_{Y}^{X} (i,\mu) = \sum_{j}e^{-\frac{1}{2\delta^{2}}(R_{ij}-\mu)^{2}}
\end{equation}

where $R_{ij}$ is the distance between central particle $i$ and neighbour $j$, $\delta$ is a fixed quantity (in our case $\delta=0.25~\mu$m), and $\mu$ is a variable parameter (we used $4.6~\mu$m$<\mu<12~\mu$m, with steps of $0.1~\mu$m). The cut-off distance from the central particle, $R_{c}$, in which this equation is calculated should include several coordination shells but is insensitive to the exact amount \cite{Cubuk2015}. In total, through varying $\mu$, we obtained a set of 75 different values for each particle, which will be referred to as \textit{features}. 

The second structure function $\Psi_{YZ}^{X} (i,\xi,\lambda,\zeta)$, related to orientational properties, is calculated as

\begin{equation}
    \label{eq:PSI_func}
    \Psi_{YZ}^{X} (i,\xi,\lambda,\zeta) = \sum_{j,k}e^{-(R_{ij}^{2}+R_{ik}^{2}+R_{jk}^{2})/\xi^{2}} \left(1+\lambda \: cos \: \theta_{ijk}\right)^{\zeta}
\end{equation}

Again, $R_{ij}$ is the distance between central particle $i$ and neighbour $j$, while $\theta_{ijk}$ is the angle that the central particle $i$ makes with its neighbours $j,k$. $\xi, \lambda, \zeta$ are variable parameters related to different aspects of the particles' local environment: $\xi$ ensures that the Gaussian exponent goes to zero as interparticle distance increases, $\lambda$ (set at either $\lambda=1$ or $\lambda=-1$) determines whether small or large bond angles are used and $\zeta$ determines or the relative importance of angular properties \cite{Schoenholz2016}. The values we used for the parameters $\xi, \lambda, \zeta$ are given in the supplemental material, giving a total of 60 features for every particle. 

\subsection{Calculating Weakness}

\begin{figure*}[t!]
 \centering
 \includegraphics[width=\textwidth]{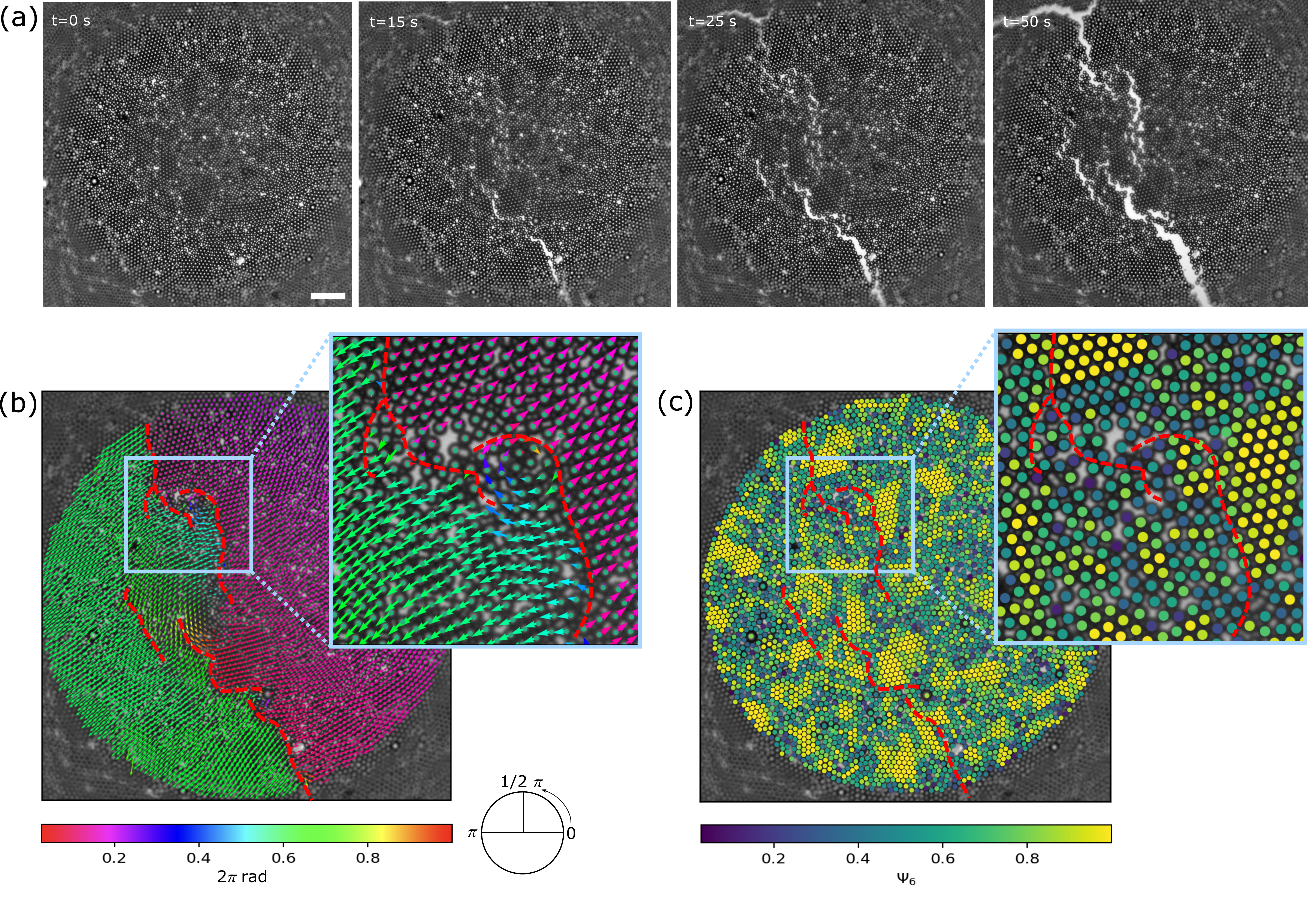}
 \caption{\textbf{(a)} Brightfield images of the colloid monolayer during a typical experiment. Scalebar represents $50~\mu$m. \textbf{(b)} Overlay of drift subtracted displacement vectors on the particle coordinates, comparing the their initial position to their position after a typical experiment (here, $t_{\rm{end}} \approx 100~$s). The vectors are coloured by their clockwise orientation, and the red striped lines show where the cracks appear in the system. \textbf{(c)} Overlay of the $\psi_6$ values on the particle initial particle positions. The red striped lines show where the cracks appear in the system.}
 \label{fig2}
\end{figure*}

We want to predict the propensity of a particle to be next to a crack line. To this end, we calculate a parameter that we will refer to as the \textit{Weakness}, which is a machine learning-generated structural indicator, calculated from the generalized description of the local environment described in Section \ref{sec:gen_parameters}. As observed in Fig. \ref{fig2}(a), crack propagation is heavily influenced by the initial crack's directionality. Therefore, we hypothesize that Weakness could identify a likely crack path in the direction of the initial crack, after its formation.

As proposed in previous work \cite{Schoenholz2016,Cubuk2015,Sharp10943} we employ one of the most straight forward machine learning algorithms: the Support Vector Machine (SVM). Support Vector Machines (SVM) are supervised classification methods, widely adopted for classification, regression and other learning tasks \cite{Chang2011}. Generally, classification algorithms have a training stage and a testing stage. During the training stage, the SVM takes as input a set of datapoints with features $x_1, x_2, ..., x_m$, providing an $m$-dimensional dataset, and for each feature a classification label (0 or 1). In our case, the datapoints are the individual particles and the features are values from Eq.~\ref{eq:G_func} and Eq.~\ref{eq:PSI_func}. The SVM algorithm then constructs and adjusts a ($m-1$)-dimensional hyperplane that separates the data into classes 0 and 1 with the highest accuracy, see Fig.~\ref{fig4}(a). Next, during the testing stage, the hyperplane is fixed and a dataset with datapoints that the algorithm has not seen before, but with the same features $x_1, x_2, ..., x_m$, is provided as input. The SVM uses previously calculated hyperplane to predict which of the two outcomes 0 or 1 is most likely for the new datapoints. 

To prevent under- and over-training we optimized the size of our dataset to approximately 12,000 randomly selected particles out of the total 100,000 particles we tracked, see Supplemental Information. The optimal ratio of particles in the dataset was found to be 45\% of particles with label 1 (crack) and 55\% with label 0 (no crack). 

% We tested the SVM using datasets of particles from a single experiment, that was excluded from the training dataset, and analyze the visual output.

We make use of the simplicity of the SVM to gain insight on the important parameters in the process. This is done by calculating the distance of datapoints from the hyperplane, which is analogous to the probability of the datapoint belonging to it's classification class. This distance has been previously used to quantify the probability for plastic rearrangements, and was termed \textit{Softness} in this context \cite{Schoenholz2016,Cubuk2015,Sharp10943}. Since in our case these labels correspond to the probability of the particle to be located on the edge of a crack, for our system we will refer to this quantity as \textit{Weakness}.

\section{Results and discussion}

\subsection{Evolution of the monolayer microstructure during fracturing}

Figure \ref{fig2}(a) shows the evolution of the colloid monolayer in a typical experiment. Cracks begin to appear shortly after expansion starts. These cracks propagate through the monolayer until they span the entire field of view, after which crack initiation ceases and crack growth proceeds through areal expansion of the already existing cracks. 

With exception of the cracks, the monolayer is not deformed, so that particles move in large rafts with the same magnitude and direction of the particles' displacement. This is visualized in Fig. \ref{fig2}(b) and inset, which shows clearly the alignment between the directional vectors of the particle movement. After fracturing, particles move away in opposite directions from the crack location, which confirms our system's suitability for studying extensional fracturing dynamics. Also, we observe that some small pockets of about $\sim 10$ particles located on the fracture line sometimes reorient themselves slightly, as seen by the rotational lines in the inset, which is a typical feature in all experiments. 

An overlay of the bond order parameter $\psi_6$ on the particle coordinates is shown in \ref{fig2}(c). The figure shows that the cracks generally propagate through regions with low $\psi_6$. This is not unexpected, since domains with $\psi_6 \rightarrow 1$ are highly ordered and densely packed so that their constituent particles are mostly surrounded by other particles fixing them in place, in contrast to more disordered domains with low $\psi_6$ where fewer interparticle bonds have to be broken for a crack to occur, thus requiring less energy.

These observations on the role of voids in crack propagation are consistent with the literature on fracturing where it was shown that voids in the crack path and near the crack tip are most prone to yielding from the crack tip-induced stresses \cite{Ravi-Chandar1984,ozawa2021rare}. The crack propagates through the material by rapid growth of these voids and their subsequent coalescence with the main crack.

\subsection{Identifying weak regions using Machine Learning}

\begin{figure}[t!]
    \centering
    \includegraphics[width=0.45\textwidth]{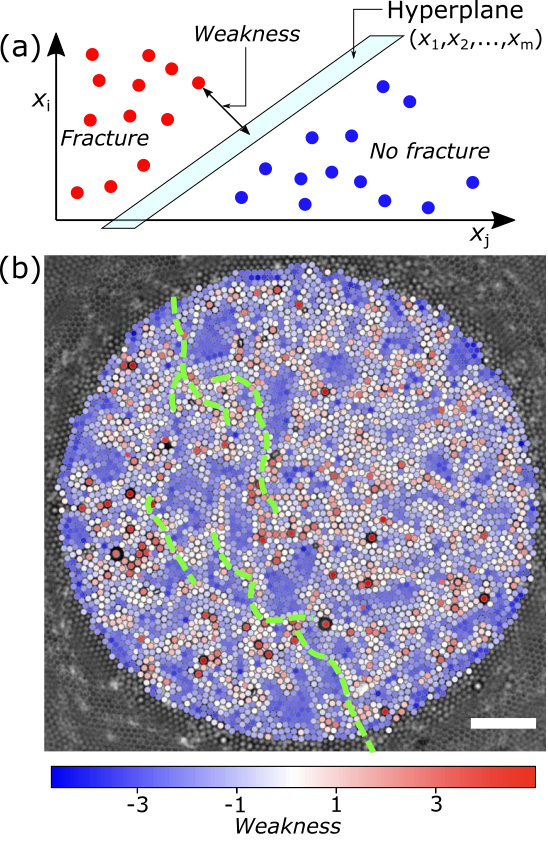}
    \caption{\textbf{(a)} Schematic of the Support Vector Machine (SVM) algorithm. We identify the distance of the particle to the hyperplane as the particle's \textit{Weakness}. \textbf{(b)} Visual overlay of the calculated Weakness values on the particles in an experimental dataset. Here, particles coloured red are located in a ``weaker'' local environment, so more prone to fracture, while blue particles are in regions that are less prone to fracture. The approximate location of the fracture in the monolayer is given by the dashed green line. Scalebar represents $50~\mu$m}
    \label{fig4}
\end{figure}

Next, we test ML algorithms for identifying regions that are prone to fracturing, and obtaining a deeper understanding of the fracturing process. First, we show how a simple ML algorithm, the SVM, can identify weak regions in the material. Then, we identify the features of the local particle environments that are most important for determining whether that region is weak. 

We show the calculated \textit{Weakness} value for each particle, obtained using a SVM, as overlay on particle coordinates in Fig.~\ref{fig4}(b). We compute the prediction accuracy by comparing the sign of the Weakness prediction, where a positive value means the particle is predicted to be on the crack line, with the classification labels for this experiment. For the experiment of Fig~\ref{fig4}(b), we find a reasonably high overall prediction accuracy of 72.8\%, which can be separated into correct predictions for particles that are not on the crack line (77.1\% accuracy) and correct predictions for particles that are on the crack line (33.0\% accuracy).

The cracks (green dashed lines in Fig. \ref{fig4}(b)) mostly appear in regions with positive Weakness values (coloured more red), so that Weakness can indeed identify a likely crack path in the crack direction. Sometimes the cracks percolate through regions of low Weakness values, which shows how the direction of the propagating crack can in some cases be dominant over the structural weakness. We hypothesize that the precise location of the propagating crack is influenced by the interplay between structural weakness and initial crack direction. This phenomenon is also observed in other experiments with a slightly different surface coverage and average ordering, see Fig. S1 in the Supplemental Information. 

\begin{figure*}[t!]
    \centering
    \includegraphics[width=0.8\textwidth]{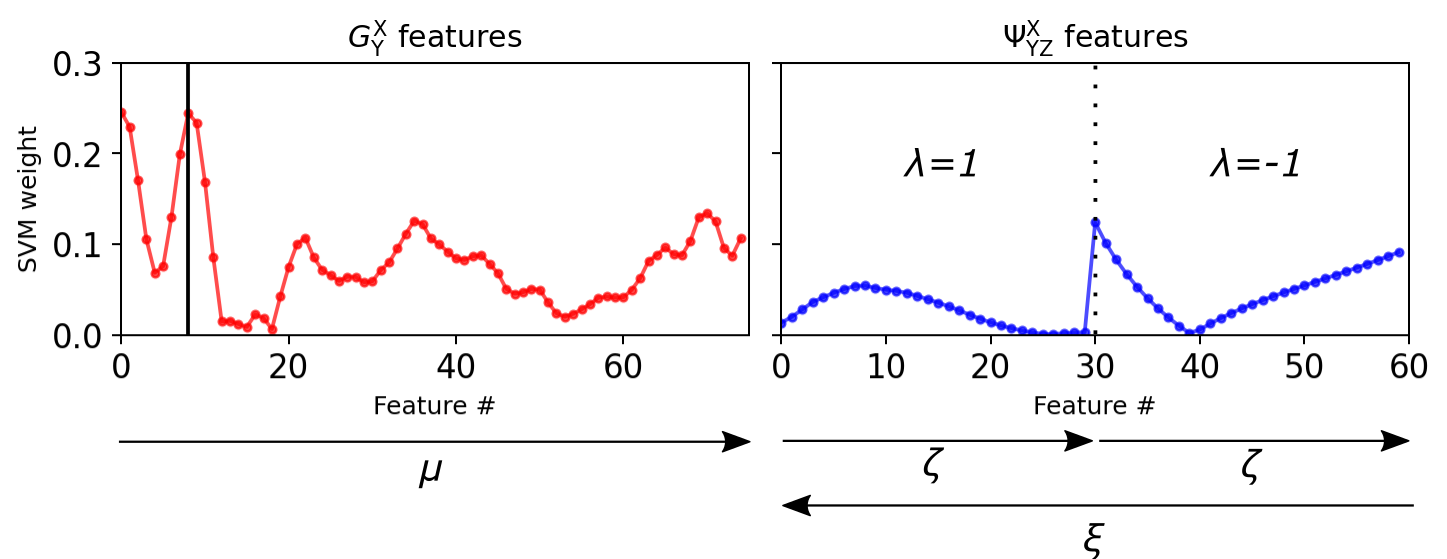}
    \caption{Weights of the SVM features, representing their relative importance. The arrows show the direction and range of the parameter variations, the numerical values are presented in Table S1 of the Supplemental Information. The vertical black line in the $G_{Y}^{X}$ plot represents feature number 8, where $\mu = 5.4~\mu$m.}
    \label{fig5}
\end{figure*}

We can characterize the alignment of cracks with regions of high Weakness values by calculating the average Weakness values of particles with label 0 and particles with label 1. We find for particles next to the crack line with label 1, in the experiment in Fig. \ref{fig4}(b), a higher average Weakness value ($-0.76$) compared to the average Weakness of all other particles ($-1.64$) with label 0, as shown in Fig. S2 in the Supplemental Information. Both average Weakness values are negative, e.g. purely from structural information the SVM method predicts it is on average still unlikely for both sets of particles to be next to a crack line. This is not unexpected, since we already observed the importance of the directionality and location of the initial crack: the crack direction can be dominant over structural weakness, meaning that the cracks sometimes propagate through regions with low Weakness values, while the cracks also propagate through only a subset of structurally weak regions in the direction of the initial crack.

The observations from the SVM algorithm output agree with the observations from structural indicators like $\psi_6$: more disordered or lower density domains are more prone to fracturing. This can be seen in Fig. \ref{fig4}(b), where particles in more ordered domains have a negative Weakness value (blue), while particles in more disordered domains are given positive Weakness values (red). 

The monolayer is slightly polydisperse and slightly larger particles are generally given a high, positive Weakness value. This is not entirely obvious, since these regions can still be dense and ordered, and those larger particles tend to have $>6$ attractive neighbours in their first coordination shell. We hypothesize that large particles might result in energetically costly point defects in the monolayer, making these points more prone to fracture \cite{Negria2015a}. Confirming would require experiments with carefully controlled defects, for instance through controlling the polydispersity of the particles, this is beyond the scope of this paper.

%The monolayer is slightly polydisperse and, interestingly, slightly larger particles are generally given a high, positive Weakness value. This is not entirely obvious, since these regions can still be dense and ordered, and those larger particles tend to have $>6$ attractive neighbours in their first coordination shell. We hypothesize that large particles might result in energetically costly point defects in the monolayer, making these points more prone to crack \cite{Negria2015a}. Confirming would require experiments with carefully controlled defects, for instance through controlling the polydispersity of the particles, which is beyond the scope of this paper.

%\mh{Since our experiments only provide a bottom-up view of the colloid monolayer, we are not able to study crack initiation, which can occur anywhere on the monolayer outside the field of view. We envision this could be solved by using high-speed 3-dimensional microscopy techniques. Furthermore, 3-dimensional structural information is most likely crucial to determining how and where the system fractures, and it might be possible to predict \textit{a priori} the initiation site and directionality of the crack, using the here proposed machine learning method.} %For instance, it is possible that the current set of structural indicators does not include some physical mechanisms that are essential to predicting where the crack occurs. Furthermore,

% It is also possible that the current set of structural indicators does not include some physical mechanisms that are essential to predicting where the crack occurs. 

\subsection{Important features in identifying weak regions}

We analyze the role of the particle features on the decision-making process of the ML algorithm by investigating more closely the mathematical formulation of the hyperplane in the SVM. The hyperplane location is determined through satisfying the equation $\bm{w^{\rm{T}}x_{i}} -b=0$, where $\bm{w^{\rm{T}}}$ is a set of weights for each feature $x_{i}$, $\bm{x_{i}}$ is a set of all features $x_{i}$ and $b$ is some offset, also referred to as the ``bias''. We note here that all our data was normalized to a domain [-1 1] before machine learning. In that case, the values obtained for $\bm{w^{\rm{T}}}$ measure the importance of the features in determining the location of the hyperplane. 

The weights for every feature number, which corresponds to a specific combination of parameters for either $G_{Y}^{X}$ or $\Psi_{YZ}^{X}$, are shown in Fig. \ref{fig5}. The arrows indicate the range and direction of the varying parameters. The figure shows that the features from the orientation-based $\Psi_{YZ}^{X}$ have a lower weight than those from the density-based $G_{Y}^{X}$. Features 30 and 60 in $\Psi_{YZ}^{X}$ have a relatively high SVM weight because for their parameter combinations the density term $e^{-(R_{ij}^{2}+R_{ik}^{2}+R_{jk}^{2})/\xi^{2}}$ is dominant. %Other variations we tested for $\zeta$ and $\xi$ did not produce significantly higher SVM weight.

These observations indicate that the local density is more important compared to the local orientational order for determining whether a domain in the material is weak. Even though the features from $\Psi_{YZ}^{X}$ also include information on the local density through the term $e^{-(R_{ij}^{2}+R_{ik}^{2}+R_{jk}^{2})/\xi^{2}}$ in Eq. \ref{eq:PSI_func}, we still make the conclusion that density is more important, since the addition of angular information in Eq. \ref{eq:PSI_func} does not in fact lead to higher SVM weight of the orientation-based features. 

The profile of the density-based features $G_{Y}^{X}$ in Fig. \ref{fig5} bears resemblance to $g(r)$, shown in Fig. \ref{fig1}(d). In fact, feature number 8, highlighted by the vertical black line and corresponding to the feature with $\mu=5.4~\mu$m, we find a peak in the SVM weights. This is striking because $\mu=5.4~\mu$m corresponds to the same location as the first coordination shell in $g(r)$. This indicates that the presence, or absence, of particles on the first coordination shell from the central particle is the most important feature in our dataset to determining Weakness.  

In Fig. \ref{fig5} the second coordination shell only corresponds to higher feature numbers approximately between 50 and 70 (where the distance between particles is $r_{ij}\sim 9.5-11.5~\mu$m). Seeing that there are multiple peaks between these two points, there are most likely recurring configurations of particles that are common in our system, providing information to determining the Weakness of those particles. Future research could make it possible to identify those shapes using shape detection algorithms.

Finally, it should be noted that we also observe high SVM weights at the lowest feature numbers in $G_{Y}^{X}$ in Fig. \ref{fig5}. We attribute these to the mis-classification of voids as particles in our experimental system. Even though our particle tracking was generally robust, each experimental dataset contained at least order $\sim 10$ misclassified particles, which is probably significant enough to show up in our results. The presence of a (misclassified) void should indeed indicate that there is a void, and thereby lead to a higher propensity to fracture.

\section{Conclusions}
In conclusion, we developed an experimental system to study a fracturing colloid monolayer on a water-air interface under isotropic dilational strain, and used structural indicators and machine learning to obtain more insight into the fracturing process. From brightfield microscopy data we obtained the particle coordinates, from which we quantified the microscructure of the monolayer through structural indicators, for instance $\psi_6$. These analyses show that cracks tend to propagate through more disordered domains. By defining and calculating the \textit{Weakness} of domains in the monolayer using machine learning, we confirmed that cracks generally propagate through structurally weak regions, but not exclusively; while crack propagation remains heavily influenced by the crack's direction and initiation site. Furthermore, we determined that local density, or the presence of voids in direct vicinity of the particle, is most important to determining whether a domain is weak to fracturing. Overall, the methodology and results presented here provide a basis for further studies into and understanding of material microstructure during fracturing. \par

\section*{Conflicts of interest}
There are no conflicts to declare.

\section*{Acknowledgements}
We thank Rafael Mayorga and Florian Meirer for discussions on crack detection and for helping to develop an algorithm to that end.

\section*{Data availability}

The datasets generated and analyzed in this study are available at the 4TU.ResearchData repository at: https://doi.org/XXXXXX.

\section*{References}

\providecommand*{\mcitethebibliography}{\thebibliography}
\csname @ifundefined\endcsname{endmcitethebibliography}
{\let\endmcitethebibliography\endthebibliography}{}

\newpage

\widetext
\begin{center}
\textsf{\textbf{\Large Supplemental Information: Microstructural features governing fracture of a two-dimensional amorphous solid identified by machine learning}}
\end{center}
%%%%%%%%%% Merge with supplemental materials %%%%%%%%%%
%%%%%%%%%% Prefix a "S" to all equations, figures, tables and reset the counter %%%%%%%%%%
\setcounter{equation}{0}
\setcounter{figure}{0}
\setcounter{table}{0}
\setcounter{section}{0}
\setcounter{page}{1}
\makeatletter
\renewcommand{\theequation}{S\arabic{equation}}
\renewcommand{\thefigure}{S\arabic{figure}}
\renewcommand{\thesection}{S\Roman{section}}
\renewcommand{\bibnumfmt}[1]{$^{\rm S#1}$}
\renewcommand{\thepage}{S\arabic{page}}

\section{Parameter choices for structure functions}

Parameter values for the structure functions $G_{Y}^{X} (i,\mu)$ and $\Psi_{YZ}^{X} (i,\xi,\lambda,\zeta)$, based on the approach used in \cite{Sharp10943}. Through these parameter variations we obtained 75 features for $G$ and 60 features for $\Psi$. We manually adjusted these to optimize quantitative output from our SVM and note that, as the paper by Behler and Parinello \cite{Behler2007} prescribes, the choice of parameters is not unique but best suits the description of the local environment in this system. 

\begin{table}[h]
\begin{center}
\caption{Feature values structure functions}
\vspace{5pt}
\label{tab:radial_func_features}
\begin{tabular}{|c|c|}
\hline
\textbf{Feature} & \textbf{Values}                      \\ \hline
$\mu \: (\mu \rm{m})$    & 4.6, 4.7, 4.8, ... , 11.8, 11.9, 12 \\ \hline \hline
$\xi (\mu \rm{m})$    & 0.368, 0.243, 0.177, ... , $e^{-\sqrt{\beta}}$, ... , 0.000492, 0.000432 \\          & where $\beta$=1, 2, 3, ... , 58, 59, 60                                                \\ \hline
$\zeta$          & 0.1, 0.2, 0.3, ... , 2.9, 3.0, 0.1, 0.2, ... , 2.8, 2.9, 3.0             \\ \hline
$\lambda$        & 1, 1, 1, ... , 1, 1, -1, -1, ... , -1, -1, -1                            \\ \hline
\end{tabular}
\end{center}
\end{table}

\section{SVM optimization}

\subsection{Choice of kernel and feature importance}

In the SVM algorithm, we used a linear kernel for the hyperplane to allow interpretation on the importance of the features. 

The hyperplane of the SVM follows the equation $\bm{w^{\rm{T}}x_{i}} -b=0$. Here, $\bm{x_{i}}$ is a set of all features $x_{i}$, $\bm{w^{\rm{T}}}$ is a set of weights for each feature $x_{i}$, and $b$ is some offset, also referred to as the ``bias''. Since $\bm{w^{\rm{T}}}$ provides the weight of a feature $x_{i}$ to the positioning of the hyperplane, this can be used as a measure of the feature importance of $x_i$. 

We obtain \textit{Weakness} by calculating the distance of datapoints from the hyperplane, which is analogous to the probability of the datapoint belonging to its classification class. 

\subsection{Dataset size and label distribution}
To prevent under- or over-fitting, we calculated the training and testing accuracy for different sized training datasets. The training accuracy was calculated through 5 fold cross validation, and we used a test set of 33,388 particles that were not used during training. The testing and training accuracy are more or less equal in the region around 12,000 datapoints, after which the testing accuracy decreases, which is a sign of overfitting \cite{Cubuk2015}. Thus, the optimal dataset size was 12,000 particles.

We also determined the optimal ratio of particles with label 1 (crack) and with label 0 (no crack). The ratio of the datapoints in the training set was varied from 10:90 to 90:10 [label 1:label 0], in steps of 5\%. Considering that equal accuracy for testing and training is optimal \cite{Cubuk2015}, we found an optimal ratio 45:55 [label 1:label 0]. 

\newpage

\begin{figure*}[t!]
    \centering
    \includegraphics[width= 0.8\textwidth]{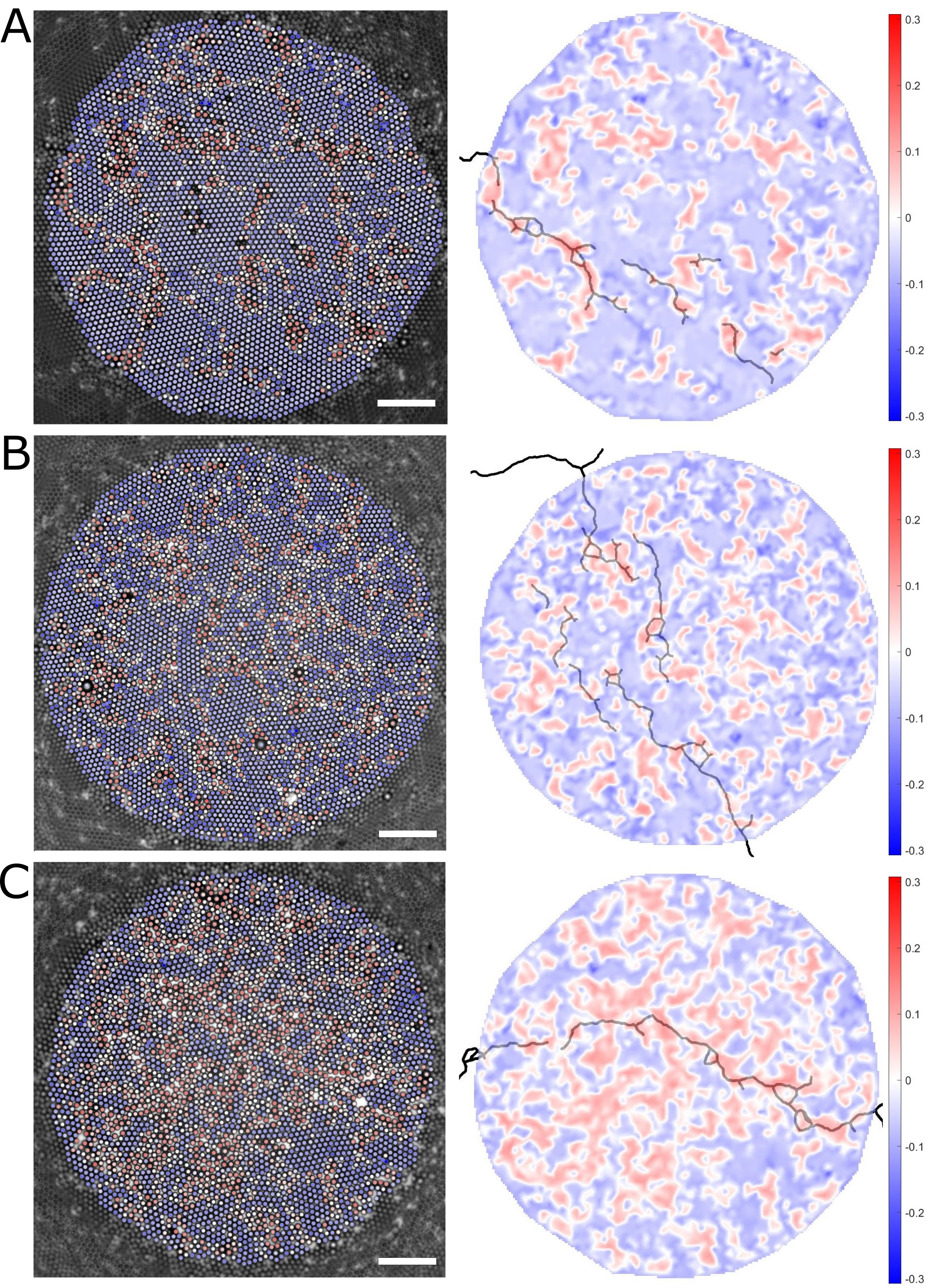}
    \caption{Machine Learning output of an SVM model using physics motivated structural indicators (left), with their corresponding surface map (right) for 3 different samples: \textbf{(a)} 3953 particles ($\varphi \approx 0.84$, prediction accuracy $=79.2\%$) \textbf{(b)} 4624 particles ($\varphi \approx 0.72$, prediction accuracy $=74.0\%$) \textbf{(c)} 3562 particles ($\varphi \approx 0.65$, prediction accuracy $=62.6\%$) The approximate location of the fracture is shown as a black line and the scalebar represent $50~\mu$m.}
    \label{fig:struct_predict}
\end{figure*}

\newpage

\begin{figure*}[t!]
    \centering
    \includegraphics[width= 0.8\textwidth]{ 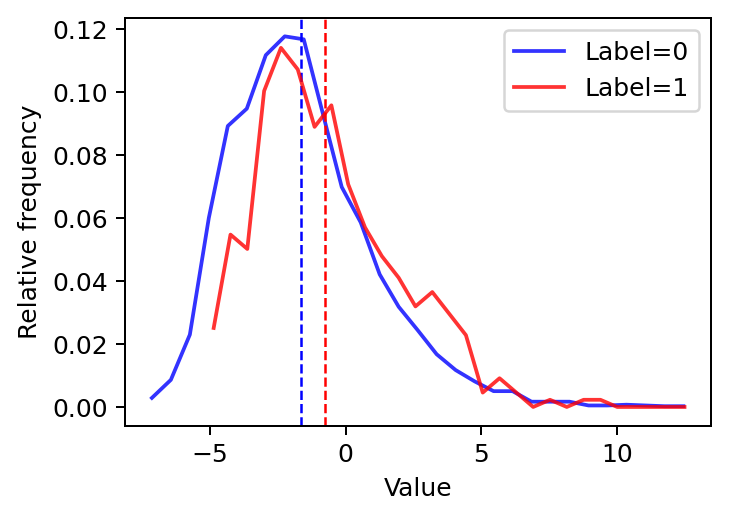}
    \caption{Distribution of predicted Weakness values. Striped lines give average Weakness for both distributions.}
    \label{fig:s2}
\end{figure*}

%%%REFERENCES%%%
\bibliography{rsc} %You need to replace "rsc" on this line with the name of your .bib file

\begin{mcitethebibliography}{30}
\providecommand*{\natexlab}[1]{#1}
\providecommand*{\mciteSetBstSublistMode}[1]{}
\providecommand*{\mciteSetBstMaxWidthForm}[2]{}
\providecommand*{\mciteBstWouldAddEndPuncttrue}
  {\def\EndOfBibitem{\unskip.}}
\providecommand*{\mciteBstWouldAddEndPunctfalse}
  {\let\EndOfBibitem\relax}
\providecommand*{\mciteSetBstMidEndSepPunct}[3]{}
\providecommand*{\mciteSetBstSublistLabelBeginEnd}[3]{}
\providecommand*{\EndOfBibitem}{}
\mciteSetBstSublistMode{f}
\mciteSetBstMaxWidthForm{subitem}
{(\emph{\alph{mcitesubitemcount}})}
\mciteSetBstSublistLabelBeginEnd{\mcitemaxwidthsubitemform\space}
{\relax}{\relax}

\bibitem[Griffith and Taylor(1921)]{Griffith1920}
A.~A. Griffith and G.~I. Taylor, \emph{Philosophical Transactions of the Royal
  Society of London. Series A, Containing Papers of a Mathematical or Physical
  Character}, 1921, \textbf{221}, 163--198\relax
\mciteBstWouldAddEndPuncttrue
\mciteSetBstMidEndSepPunct{\mcitedefaultmidpunct}
{\mcitedefaultendpunct}{\mcitedefaultseppunct}\relax
\EndOfBibitem
\bibitem[Irwin(1957)]{Irwin1957}
G.~R. Irwin, \emph{J. Appl. Mech.}, 1957, \textbf{24}, 361--364\relax
\mciteBstWouldAddEndPuncttrue
\mciteSetBstMidEndSepPunct{\mcitedefaultmidpunct}
{\mcitedefaultendpunct}{\mcitedefaultseppunct}\relax
\EndOfBibitem
\bibitem[Williams(1961)]{Williams1961}
M.~L. Williams, \emph{J. Appl. Mech.}, 1961, \textbf{28}, 78--82\relax
\mciteBstWouldAddEndPuncttrue
\mciteSetBstMidEndSepPunct{\mcitedefaultmidpunct}
{\mcitedefaultendpunct}{\mcitedefaultseppunct}\relax
\EndOfBibitem
\bibitem[Westergaard(2021)]{Westergaard1939}
H.~M. Westergaard, \emph{J. Appl. Mech.}, 2021, \textbf{6}, A49--A53\relax
\mciteBstWouldAddEndPuncttrue
\mciteSetBstMidEndSepPunct{\mcitedefaultmidpunct}
{\mcitedefaultendpunct}{\mcitedefaultseppunct}\relax
\EndOfBibitem
\bibitem[Rountree \emph{et~al.}(2002)Rountree, Kalia, Lidorikis, Nakano,
  Van~Brutzel, and Vashishta]{Rountree2002}
C.~L. Rountree, R.~K. Kalia, E.~Lidorikis, A.~Nakano, L.~Van~Brutzel and
  P.~Vashishta, \emph{Annu. Rev. Mater. Res.}, 2002, \textbf{32},
  377--400\relax
\mciteBstWouldAddEndPuncttrue
\mciteSetBstMidEndSepPunct{\mcitedefaultmidpunct}
{\mcitedefaultendpunct}{\mcitedefaultseppunct}\relax
\EndOfBibitem
\bibitem[Rozen-Levy \emph{et~al.}(2020)Rozen-Levy, Kolinski, Cohen, and
  Fineberg]{Rozen-Levy2020}
L.~Rozen-Levy, J.~M. Kolinski, G.~Cohen and J.~Fineberg, \emph{Phys. Rev.
  Lett.}, 2020, \textbf{125}, 175501\relax
\mciteBstWouldAddEndPuncttrue
\mciteSetBstMidEndSepPunct{\mcitedefaultmidpunct}
{\mcitedefaultendpunct}{\mcitedefaultseppunct}\relax
\EndOfBibitem
\bibitem[Ozawa \emph{et~al.}(2022)Ozawa, Berthier, Biroli, and
  Tarjus]{ozawa2021rare}
M.~Ozawa, L.~Berthier, G.~Biroli and G.~Tarjus, \emph{Phys. Rev. Res.}, 2022,
  \textbf{4}, 023227\relax
\mciteBstWouldAddEndPuncttrue
\mciteSetBstMidEndSepPunct{\mcitedefaultmidpunct}
{\mcitedefaultendpunct}{\mcitedefaultseppunct}\relax
\EndOfBibitem
\bibitem[Kalia \emph{et~al.}(1997)Kalia, Nakano, Omeltchenko, Tsuruta, and
  Vashishta]{Kalia1997}
R.~K. Kalia, A.~Nakano, A.~Omeltchenko, K.~Tsuruta and P.~Vashishta,
  \emph{Phys. Rev. Lett.}, 1997, \textbf{78}, 2144--2147\relax
\mciteBstWouldAddEndPuncttrue
\mciteSetBstMidEndSepPunct{\mcitedefaultmidpunct}
{\mcitedefaultendpunct}{\mcitedefaultseppunct}\relax
\EndOfBibitem
\bibitem[Ravi-Chandar and Knauss(1984)]{Ravi-Chandar1984}
K.~Ravi-Chandar and W.~G. Knauss, \emph{International Journal of Fracture},
  1984, \textbf{26}, 141--154\relax
\mciteBstWouldAddEndPuncttrue
\mciteSetBstMidEndSepPunct{\mcitedefaultmidpunct}
{\mcitedefaultendpunct}{\mcitedefaultseppunct}\relax
\EndOfBibitem
\bibitem[Pollard and Fielding(2022)]{Pollard2022}
J.~Pollard and S.~M. Fielding, \emph{Phys. Rev. Res.}, 2022, \textbf{4},
  043037\relax
\mciteBstWouldAddEndPuncttrue
\mciteSetBstMidEndSepPunct{\mcitedefaultmidpunct}
{\mcitedefaultendpunct}{\mcitedefaultseppunct}\relax
\EndOfBibitem
\bibitem[Aime \emph{et~al.}(2018)Aime, Ramos, and Cipelletti]{Aime2018}
S.~Aime, L.~Ramos and L.~Cipelletti, \emph{Proc. Natl. Acad. Sci. U. S. A.},
  2018, \textbf{115}, 3587--3592\relax
\mciteBstWouldAddEndPuncttrue
\mciteSetBstMidEndSepPunct{\mcitedefaultmidpunct}
{\mcitedefaultendpunct}{\mcitedefaultseppunct}\relax
\EndOfBibitem
\bibitem[Giorgiutti-Dauphiné and Pauchard(2016)]{Giorgiutti2016}
F.~Giorgiutti-Dauphiné and L.~Pauchard, \emph{J. Appl. Phys.}, 2016,
  \textbf{120}, 065107\relax
\mciteBstWouldAddEndPuncttrue
\mciteSetBstMidEndSepPunct{\mcitedefaultmidpunct}
{\mcitedefaultendpunct}{\mcitedefaultseppunct}\relax
\EndOfBibitem
\bibitem[Floch-Fouéré \emph{et~al.}(2019)Floch-Fouéré, Lanotte, Jeantet,
  and Pauchard]{Floch2019}
C.~L. Floch-Fouéré, L.~Lanotte, R.~Jeantet and L.~Pauchard, \emph{Soft
  Matter}, 2019, \textbf{15}, 6190--6199\relax
\mciteBstWouldAddEndPuncttrue
\mciteSetBstMidEndSepPunct{\mcitedefaultmidpunct}
{\mcitedefaultendpunct}{\mcitedefaultseppunct}\relax
\EndOfBibitem
\bibitem[Buttinoni \emph{et~al.}(2017)Buttinoni, Steinacher, Spanke, Pokki,
  Bahmann, Nelson, Foffi, and Isa]{Buttinoni2017a}
I.~Buttinoni, M.~Steinacher, H.~T. Spanke, J.~Pokki, S.~Bahmann, B.~Nelson,
  G.~Foffi and L.~Isa, \emph{Phys. Rev. E}, 2017, \textbf{95}, 1--11\relax
\mciteBstWouldAddEndPuncttrue
\mciteSetBstMidEndSepPunct{\mcitedefaultmidpunct}
{\mcitedefaultendpunct}{\mcitedefaultseppunct}\relax
\EndOfBibitem
\bibitem[Galloway \emph{et~al.}(2020)Galloway, Ma, Keim, Jerolmack, Yodh, and
  Arratia]{Galloway2020}
K.~L. Galloway, X.~Ma, N.~C. Keim, D.~J. Jerolmack, A.~G. Yodh and P.~E.
  Arratia, \emph{Proc. Natl. Acad. Sci. U. S. A.}, 2020, \textbf{117},
  11887--11893\relax
\mciteBstWouldAddEndPuncttrue
\mciteSetBstMidEndSepPunct{\mcitedefaultmidpunct}
{\mcitedefaultendpunct}{\mcitedefaultseppunct}\relax
\EndOfBibitem
\bibitem[Buttinoni \emph{et~al.}(2017)Buttinoni, Cha, Lin, Job, Daraio, and
  Isa]{Buttinoni2017}
I.~Buttinoni, J.~Cha, W.~H. Lin, S.~Job, C.~Daraio and L.~Isa, \emph{Proc.
  Natl. Acad. Sci. U. S. A.}, 2017, \textbf{114}, 12150--12155\relax
\mciteBstWouldAddEndPuncttrue
\mciteSetBstMidEndSepPunct{\mcitedefaultmidpunct}
{\mcitedefaultendpunct}{\mcitedefaultseppunct}\relax
\EndOfBibitem
\bibitem[Richard \emph{et~al.}(2020)Richard, Ozawa, Patinet, Stanifer, Shang,
  Ridout, Xu, Zhang, Morse, Barrat, Berthier, Falk, Guan, Liu, Martens, Sastry,
  Vandembroucq, Lerner, and Manning]{Richard2020}
D.~Richard, M.~Ozawa, S.~Patinet, E.~Stanifer, B.~Shang, S.~A. Ridout, B.~Xu,
  G.~Zhang, P.~K. Morse, J.-L. Barrat, L.~Berthier, M.~L. Falk, P.~Guan, A.~J.
  Liu, K.~Martens, S.~Sastry, D.~Vandembroucq, E.~Lerner and M.~L. Manning,
  \emph{Phys. Rev. Materials}, 2020, \textbf{4}, 113609\relax
\mciteBstWouldAddEndPuncttrue
\mciteSetBstMidEndSepPunct{\mcitedefaultmidpunct}
{\mcitedefaultendpunct}{\mcitedefaultseppunct}\relax
\EndOfBibitem
\bibitem[Schoenholz \emph{et~al.}(2016)Schoenholz, Cubuk, Sussman, Kaxiras, and
  Liu]{Schoenholz2016}
S.~S. Schoenholz, E.~D. Cubuk, D.~M. Sussman, E.~Kaxiras and A.~J. Liu,
  \emph{Nature Phys}, 2016, \textbf{12}, 469--471\relax
\mciteBstWouldAddEndPuncttrue
\mciteSetBstMidEndSepPunct{\mcitedefaultmidpunct}
{\mcitedefaultendpunct}{\mcitedefaultseppunct}\relax
\EndOfBibitem
\bibitem[Cubuk \emph{et~al.}(2015)Cubuk, Schoenholz, Rieser, Malone, Rottler,
  Durian, Kaxiras, and Liu]{Cubuk2015}
E.~D. Cubuk, S.~S. Schoenholz, J.~M. Rieser, B.~D. Malone, J.~Rottler, D.~J.
  Durian, E.~Kaxiras and A.~J. Liu, \emph{Phys. Rev. Lett.}, 2015,
  \textbf{114}, 108001\relax
\mciteBstWouldAddEndPuncttrue
\mciteSetBstMidEndSepPunct{\mcitedefaultmidpunct}
{\mcitedefaultendpunct}{\mcitedefaultseppunct}\relax
\EndOfBibitem
\bibitem[Sharp \emph{et~al.}(2018)Sharp, Thomas, Cubuk, Schoenholz, Srolovitz,
  and Liu]{Sharp10943}
T.~A. Sharp, S.~L. Thomas, E.~D. Cubuk, S.~S. Schoenholz, D.~J. Srolovitz and
  A.~J. Liu, \emph{Proc. Natl. Acad. Sci. U. S. A.}, 2018, \textbf{115},
  10943--10947\relax
\mciteBstWouldAddEndPuncttrue
\mciteSetBstMidEndSepPunct{\mcitedefaultmidpunct}
{\mcitedefaultendpunct}{\mcitedefaultseppunct}\relax
\EndOfBibitem
\bibitem[Huerre \emph{et~al.}(2018)Huerre, Cacho-Nerin, Poulichet, Udoh,
  Corato, and Garbin]{Huerre2018}
A.~Huerre, F.~Cacho-Nerin, V.~Poulichet, C.~E. Udoh, M.~D. Corato and
  V.~Garbin, \emph{Langmuir}, 2018, \textbf{34}, 1020--1028\relax
\mciteBstWouldAddEndPuncttrue
\mciteSetBstMidEndSepPunct{\mcitedefaultmidpunct}
{\mcitedefaultendpunct}{\mcitedefaultseppunct}\relax
\EndOfBibitem
\bibitem[Poulichet and Garbin(2015)]{Poulichet2015_langmuir}
V.~Poulichet and V.~Garbin, \emph{Langmuir}, 2015, \textbf{31},
  12035--12042\relax
\mciteBstWouldAddEndPuncttrue
\mciteSetBstMidEndSepPunct{\mcitedefaultmidpunct}
{\mcitedefaultendpunct}{\mcitedefaultseppunct}\relax
\EndOfBibitem
\bibitem[Allan \emph{et~al.}(2019)Allan, van~der Wel, Keim, Caswell, Wieker,
  Verweij, Reid, Thierry, Grueter, Ramos, apiszcz, zoeith, Perry, Boulogne,
  Sinha, pfigliozzi, Bruot, Uieda, Katins, Mary, and
  Ahmadia]{dan_allan_2019_3492186}
D.~Allan, C.~van~der Wel, N.~Keim, T.~A. Caswell, D.~Wieker, R.~Verweij,
  C.~Reid, Thierry, L.~Grueter, K.~Ramos, apiszcz, zoeith, R.~W. Perry,
  F.~Boulogne, P.~Sinha, pfigliozzi, N.~Bruot, L.~Uieda, J.~Katins, H.~Mary and
  A.~Ahmadia, \emph{soft-matter/trackpy: Trackpy v0.4.2}, 2019,
  \url{https://doi.org/10.5281/zenodo.3492186}\relax
\mciteBstWouldAddEndPuncttrue
\mciteSetBstMidEndSepPunct{\mcitedefaultmidpunct}
{\mcitedefaultendpunct}{\mcitedefaultseppunct}\relax
\EndOfBibitem
\bibitem[Crocker and Grier(1996)]{CROCKER1996298}
J.~C. Crocker and D.~G. Grier, \emph{J. Colloid Interface Sci.}, 1996,
  \textbf{179}, 298 -- 310\relax
\mciteBstWouldAddEndPuncttrue
\mciteSetBstMidEndSepPunct{\mcitedefaultmidpunct}
{\mcitedefaultendpunct}{\mcitedefaultseppunct}\relax
\EndOfBibitem
\bibitem[Mayorga-Gonz{\'a}lez \emph{et~al.}(2021)Mayorga-Gonz{\'a}lez,
  Rivera-Torrente, Nikolopoulos, Bossers, Valadian, Yus, Seoane, Weckhuysen,
  and Meirer]{mayorga2021visualizing}
R.~Mayorga-Gonz{\'a}lez, M.~Rivera-Torrente, N.~Nikolopoulos, K.~W. Bossers,
  R.~Valadian, J.~Yus, B.~Seoane, B.~M. Weckhuysen and F.~Meirer, \emph{Chem.
  Sci.}, 2021, \textbf{12}, 8458--8467\relax
\mciteBstWouldAddEndPuncttrue
\mciteSetBstMidEndSepPunct{\mcitedefaultmidpunct}
{\mcitedefaultendpunct}{\mcitedefaultseppunct}\relax
\EndOfBibitem
\bibitem[Schwenke \emph{et~al.}(2014)Schwenke, Isa, and Del~Gado]{Schwenke2014}
K.~Schwenke, L.~Isa and E.~Del~Gado, \emph{Langmuir}, 2014, \textbf{30},
  3069--3074\relax
\mciteBstWouldAddEndPuncttrue
\mciteSetBstMidEndSepPunct{\mcitedefaultmidpunct}
{\mcitedefaultendpunct}{\mcitedefaultseppunct}\relax
\EndOfBibitem
\bibitem[Keim and Arratia(2014)]{Keim2014}
N.~C. Keim and P.~E. Arratia, \emph{Phys. Rev. Lett.}, 2014, \textbf{112},
  028302\relax
\mciteBstWouldAddEndPuncttrue
\mciteSetBstMidEndSepPunct{\mcitedefaultmidpunct}
{\mcitedefaultendpunct}{\mcitedefaultseppunct}\relax
\EndOfBibitem
\bibitem[Behler and Parrinello(2007)]{Behler2007}
J.~Behler and M.~Parrinello, \emph{Phys. Rev. Lett.}, 2007, \textbf{98},
  146401\relax
\mciteBstWouldAddEndPuncttrue
\mciteSetBstMidEndSepPunct{\mcitedefaultmidpunct}
{\mcitedefaultendpunct}{\mcitedefaultseppunct}\relax
\EndOfBibitem
\bibitem[Chang and Lin(2011)]{Chang2011}
C.-C. Chang and C.-J. Lin, \emph{ACM Trans. Intell. Syst. Technol.}, 2011,
  \textbf{2}, 27\relax
\mciteBstWouldAddEndPuncttrue
\mciteSetBstMidEndSepPunct{\mcitedefaultmidpunct}
{\mcitedefaultendpunct}{\mcitedefaultseppunct}\relax
\EndOfBibitem
\bibitem[Negria \emph{et~al.}(2015)Negria, Sellerioa, Zapperia, and
  Miguele]{Negria2015a}
C.~Negria, A.~L. Sellerioa, S.~Zapperia and M.~C. Miguele, \emph{Proc. Natl.
  Acad. Sci. U. S. A.}, 2015, \textbf{112}, 14545--14550\relax
\mciteBstWouldAddEndPuncttrue
\mciteSetBstMidEndSepPunct{\mcitedefaultmidpunct}
{\mcitedefaultendpunct}{\mcitedefaultseppunct}\relax
\EndOfBibitem
\end{mcitethebibliography}
\bibliographystyle{unsrt} %the RSC's .bst file

\end{document}